\documentclass[reprint,pra,aps,nofootinbib,longbibliography,superscriptaddress]{revtex4-2}

\usepackage{amsmath,amssymb,amstext}
\usepackage{mathtools}
\usepackage[pdftex]{graphicx}
\usepackage{tabularx}
\usepackage{bm}
\usepackage{mathrsfs}
\usepackage{dsfont}
\usepackage{fancyhdr}
\usepackage{booktabs}
\usepackage{longtable}
\usepackage{multirow}
\usepackage[normalem]{ulem}
\usepackage{xcolor}
\usepackage{float}

\usepackage[pdftex,pagebackref=false]{hyperref}
\hypersetup{
    unicode=false,          
    pdffitwindow=false,     
    pdfstartview={FitH},    
    pdfnewwindow=true,      
    colorlinks=true,        
    linkcolor=blue,         
    citecolor=green,        
    filecolor=magenta,      
    urlcolor=cyan           
}

\usepackage{braket}
\usepackage{amsthm}
\usepackage{bbold}
\usepackage{comment}

\newcommand{\mf}[1]{\mathsf{#1}}
\newcommand{\Tr}{\operatorname{Tr}}

\newcommand{\dd}{\mathrm{d}}
\newcommand{\ii}{\mathrm{i}}

\newcommand{\tbb}[1]{\textcolor{blue}{#1}}

\allowdisplaybreaks

\begin{document}



\title{
Spacetime Emergence in the Memory of a Learning Agent
}





\author{Philip A. LeMaitre}
\email{Philip.Lemaitre@uibk.ac.at}
\affiliation{University of Innsbruck, Institute for Theoretical Physics, Technikerstrasse 21a, A-6020 Innsbruck, Austria}
\author{T. Rick Perche}
\email{rick.perche@su.se}
\affiliation{Nordita, KTH Royal Institute of Technology and Stockholm University,  Hannes Alfvéns väg 12, 23, SE-106 91 Stockholm, Sweden}

\date{\today}

\begin{abstract}


We show that the causal structure of spacetime can emerge within the memory of a learning agent performing a quantum information task. We train a model-based reinforcement-learning agent to maximize the quantum mutual information between localized quantum probes interacting with a quantum field in Minkowski spacetime, without directly providing it with information regarding the spacetime metric or causal structure. In learning to accomplish this task, the reinforcement-learning agent constructs an internal predictive model whose planner develops the characteristic null structure of spacetime, allowing us to reconstruct the causal structure of the Minkowski metric with high accuracy. Within this task, spacetime geometry is obtained from an agent's successful predictions and actions, providing a first step toward showing that spacetime concepts emerge from operational quantum experience.
\end{abstract}

\maketitle

\section{Introduction} \label{Section:Introduction}

General relativity is our most successful description of spacetime and has passed a remarkably broad range of experimental tests to date~\cite{Will2014}, yet the geometry described by the theory --- commonly represented by a Lorentzian metric tensor, which is an object encoding spacetime distances, angles, and causal relationships~\cite{Wald1984,EhlersPiraniSchild2012} --- is never accessed directly in experiment but is instead inferred from the records left by physical systems interacting with matter and fields~\cite{EhlersPiraniSchild2012,Basri1965}. Whether through families of causal trajectories or the responses of localized detectors, every observation provides only finite operational information from which the underlying geometry must be reconstructed~\cite{Synge1960,EhlersPiraniSchild2012,Perlick2008,HawkingKingMcCarthy1976,Malament1977,SaravaniAslanbeigiKempf2016,PercheMartinMartinez2022Geometry,Haag}. This operational viewpoint shifts the emphasis from geometry itself to the physical records encoding that geometry, motivating broader questions that lie at the heart of spacetime reconstruction: what kind of records are sufficient for an observer to recover the structure of the underlying spacetime and to what extent do they preserve the geometric information of the spacetime that generated them? 

Multiple approaches have been developed to try and answer these questions, with quantum-information-based schemes being of particular interest with regards to what will be considered in the current work (introduced further down). For instance, Ref. ~\cite{CaoCarrollMichalakis2017} reconstructs spatial geometry from the entanglement structure of a Hilbert space factorization by using mutual information to define effective distances before fitting an emergent geometry. Ref. ~\cite{Kempf2021CorrelationDistance} advocates for replacing spacetime distance by correlation strength, treating abstract correlators as the primary structure from which spacetime and fields may arise as approximate descriptions. Additionally, Ref.~\cite{FranzmannJovancicLawson2023} analyzes how relative distance can be associated with mutual information in entanglement-based emergent-spacetime scenarios while showing that simple entangled systems already contain ambiguities in such reconstructions. 

Relativistic quantum information (RQI) arguably provides the most natural framework in which to address this question. In particular, Unruh–DeWitt (UDW) detectors~\cite{Unruh1976,DeWitt1979} furnish a simple model of localized quantum systems interacting with quantum fields ~\cite{FewsterVerch2020,GrimmerTorresMartinMartinez2021} while remaining sufficiently tractable that the resulting detector states can be analyzed using the tools of quantum information theory~\cite{LoukoCurvedSpacetimes,Martin-Martinez_2015,Martin-Martinez_Perche_Torres_2021}. It is through these interactions that geometric and causal relationships as well as field correlations become encoded into finite-dimensional quantum states whose operational properties can be studied directly. UDW systems have consequently become one of the principal tools ~\cite{CrispinoHiguchiMatsas2008} for investigating how spacetime structure is reflected in quantum information, leading to a variety of protocols for reconstructing geometric quantities from detector data~\cite{MartinMartinezSmithTerno2016SpacetimeStructure,PercheMartinMartinez2022Geometry,SaravaniAslanbeigiKempf2016}. 

The perspective adopted here is somewhat different to the above mentioned works in that we ask whether a learning agent interacting only with operational quantum data sourced from quantum systems embedded in a spacetime (and with no \emph{a priori} knowledge of spacetime), can itself acquire an internal representation that reflects the geometry of the spacetime generating those observations. The question is therefore whether successful interaction with a relativistic quantum environment naturally gives rise to representations from which that geometry can subsequently be extracted. Viewed in this way, a learning agent serves as a concrete operational observer whose internal memory is shaped by experience. More concretely, our setting considers model-based reinforcement-learning (RL) agents acting on an environment defined by the outcomes of a quantum information protocol performed in spacetime. RL converts the structure of that environment into a memory of actions that we can attempt to query to ascertain whether it contains a recoverable metric structure.

Our application of machine learning in pursuit of answering this question is a somewhat recent development with only a relatively small number of works employing the same strategy, and fewer still using RL. For example, the authors of Ref. ~\cite{YouYangQi2018} train a model on entanglement features so that the learned tensor-network geometry reproduces the entanglement structure of a quantum many-body state; the authors of Refs.~\cite{HashimotoSugishitaTanakaTomiya2018AdSCFT,HashimotoSugishitaTanakaTomiya2018QCD} use deep learning in AdS/CFT-inspired settings to infer emergent bulk metrics from field-theoretic or lattice data, and more recently used interpretable neural architectures to reconstruct gravitational metrics from linear-response data~\cite{Hashimoto_Matsuo_Murata_Ogiwara_Takeda_2025}; recent holographic reconstruction work uses entanglement-entropy data as input to a machine-learning inverse problem for the bulk metric~\cite{AhnJeongKimYun2024}; and in the RQI sphere, Ref. ~\cite{Grimmer_Melgarejo-Lermas_Polo-Gomez_Martin-Martinez_2023} uses deep learning to distinguish field states, detect field thermalization, and sense spacetime boundaries. Work on quantum-enhanced agents and causal machine learning in quantum settings comes closer still by treating agents as active participants whose actions and records mediate what can be learned about a quantum environment~\cite{shrapnel2020quantum}. Our approach combines this active participant idea with RL in an RQI setting in a way that answers the original questions, but also distinguishes itself from these other works by drawing attention to the agentic aspect of learning new concepts from acting. In this respect, the problem is also related to work on intuitive physics and physical concept learning, where predictive models acquire internal representations of physical regularities from experience rather than having access to an explicit symbolic theory~\cite{Piloto2022}.

To investigate these questions, we construct a detector-generated environment in which one Gaussian-smeared UDW detector remains fixed while a second is translated across a two-dimensional grid of spacetime smearing centres $\mf{x}=(x,t)$. Their interaction with a real massless scalar quantum field in $(3+1)$-dimensional Minkowski spacetime, holding two spatial dimensions fixed, generates a transformed quantum mutual-information objective that provides the sole reward to a model-based RL agent with no initial conception of spacetime. The underlying spacetime structure is implicitly encoded in this reward and so the RL agent must learn a predictive representation of the environment --- and thus the spacetime structure --- through interaction with it. We then ask whether the memory supporting its successful behaviour retains enough of the structure underlying the operational quantum information for the spacetime geometry to be reconstructed from it. For this we apply a mask to the learned planner representation that minimally identifies lightcone-like branches, to which we fit a quadratic form whose coefficients yield the metric tensor components $g_{tt}\simeq-0.99988$ and $g_{tx}\simeq 7.79\times10^{-8}$ (after fixing the conformal freedom by setting $g_{xx}=1$) with null slopes $dx/dt\simeq\pm0.99994$, in close agreement with the corresponding Minkowski values. Since null structure determines the Lorentzian metric only up to a conformal factor~\cite{HawkingKingMcCarthy1976,Malament1977,EhlersPiraniSchild2012}, this demonstrates that an internal representation learned solely for prediction and action can preserve sufficient operational information for the local conformal geometry of the spacetime to then be recovered.

The remainder of the paper is organized as follows. Section~\ref{Section:UdW} establishes the basics of the UDW model and the framework of Algebraic Quantum Field Theory (AQFT), then derives the relevant quantities for the detector-generated reward landscape. Section~\ref{Section:RL} briefly introduces the basics of RL and details the architecture of the learning agent. The details of the learning task and RL agent training procedure are then given in Section ~\ref{Section:LearningTask}, and subsequently Section~\ref{Section:AgentMemory} presents the reconstruction of the spacetime metric from the learned representation in the RL agent's memory. Finally, Section~\ref{Section:Discussion} discusses the implications, limitations, and possible extensions of the framework. The code used to obtain the results can be found in an accompanying GitHub repository ~\cite{SpacetimeML_Github}. 

\section{Unruh-DeWitt Model} \label{Section:UdW}

The UDW model bridges quantum information with relativity in a concrete way such that a learning task based on quantum information may be phrased in a spacetime setting. From the choice of a spacetime region to couple the detector to and a field state, one can determine the final finite-dimensional detector state whose correlations are then used to construct the reward landscape. The RL agent has access only to this resulting reward, while the spacetime geometry and field-theoretic quantities from which it is generated remain directly inaccessible. Each point in the environment therefore corresponds to a possible spacetime smearing centre for detector 2, with the value assigned to that point determined by the correlations induced through its interaction with the field.

UDW detectors were introduced as probes of quantum fields and observer-dependent particle content~\cite{Unruh1976,DeWitt1979,Takagi}.  Their appeal in relativistic quantum information comes from the fact that a finite-dimensional quantum system can be coupled locally to a quantum field while keeping the operational output on the level of detector states. This has made the model remarkably useful for studying the transition probabilities, field noise, signalling, covariance, and the dependence of detector response on switching and localization~\cite{LoukoCurvedSpacetimes,Schlicht_2003,Martin-Martinez_2015,Martin-Martinez_Perche_Torres_2021,deRamon_Papageorgiou_Martin-Martinez_2023}; as well as providing a means for operationalizing spacetime properties into quantum information and computation resources for use in entanglement/contextuality harvesting ~\cite{Pozas-Kerstjens_Martin-Martinez_2015,Pozas-Kerstjens_Martin-Martinez_2016,LeMaitre2026HarvestingContext} and relativistic quantum computing~\cite{Martin-Martinez_Aasen_Kempf_2013,Bruschi_Dragan_Lee_Fuentes_Louko_2013,Martín-Martínez_Sutherland_2014,AsplingLawler2024,LeMaitre_Perche_Krumm_Briegel_2024}. Work also shows how detector models capture fundamental features of the light-matter interaction~\cite{lopp2021quantum} and how they extend to linearized-gravity settings~\cite{Perche_Ragula_Martin-Martinez_2023}. 

For the purpose of this manuscript, we consider two qubit UDW detectors interacting with a real massless scalar quantum field $\hat{\phi}(\mf x)$ in 3+1-dimensional Minkowski spacetime, whose dynamics are prescribed by the Klein-Gordon equation
\begin{align}
\nabla^\mu\nabla_\mu \hat \phi(\mf x) = 0\,,
\end{align}
and can be decomposed as
\begin{align}
\hat{\phi}(\mf x) = \int \frac{\mathrm{d}^3\bm k}{\sqrt{2(2\pi^3)|\bm k|}} \left(\hat{a}_{\bm k}^{\dagger} e^{i(|\bm k|t-\bm k \cdot \bm x)}+\hat{a}_{\bm k}e^{-i(|\bm k|t-\bm k \cdot \bm x)}\right)\,,
\end{align}
where $\hat{a}_{\bm k}, \hat{a}^{\dagger}_{\bm k}$ are the field ladder operators satisfying the commutation relation $[\hat{a}_{\bm k}, \hat{a}^{\dagger}_{\bm k'}] = \delta^3(\bm k - \bm k')$.

The detectors are modeled as qubits with Hilbert spaces $\mathcal{H}_1\cong \mathcal{H}_2 \cong \mathbb{C}^2$, each moving along comoving inertial timelike worldlines $\mf z_d^\mu(t) = (t, \bm x_d)\ , d \in\{1,2\}$, where $t$ is an affine inertial time parameter that also represents the detectors' proper time. Along each worldline, each qubit has internal dynamics governed by a free Hamiltonian $\hat{H}_{d} = \frac{\Omega_d}{2}(\mathds{1} - \hat{\sigma}_z^{(d)})$, where $\Omega_d$ denotes the energy gap between the ground and excited states $\ket{0_d}$ and $\ket{1_d}$, respectively, so that $\hat{\sigma}_z^{(d)} = \ket{0_d}\!\!\bra{0_d} - \ket{1_d}\!\!\bra{1_d}$. Additionally, each qubit interacts with a massless scalar quantum field $\hat{\phi}(\mf x)$ locally around their trajectories. In the local description of field theory (see e.g. Ref. ~\cite{Haag}), the scalar field $\hat{\phi}(\mf x)$ satisfies the commutation relation 
\begin{align}\label{Equation:CovCanCom}
\left[\hat{\phi}(\mf x), \hat{\phi}(\mf x')\right] = iE(\mf x, \mf x')\mathds{1}
\end{align}
where $E(\mf x, \mf x') = G_R(\mf x, \mf x')-G_A(\mf x, \mf x')$ is called the causal propagator, with $G_R(\mf x, \mf x')$ and $G_A(\mf x, \mf x')$ being the retarded and advanced propagators of the field, respectively. The causal propagator is related to the Wightman function of the field $W(\mf x,\mf x') = \langle \hat{\phi}(\mf x)\hat{\phi}(\mf x')\rangle_{\hat{\rho}_\phi}$, through the decomposition 
\begin{align}
W(\mf x, \mf x') &= \frac{1}{2}\left\langle \left \{\hat{\phi}(\mf x), \hat{\phi}(\mf x')\right\}\right\rangle_{\hat{\rho}_\phi}\!\!\!+\frac{1}{2}\left\langle \left[\hat{\phi}(\mf x), \hat{\phi}(\mf x')\right]\right\rangle_{\hat{\rho}_\phi}\nonumber \\[1ex]
&= \frac{1}{2}H(\mf x, \mf x') +\frac{i}{2}E(\mf x, \mf x')
\end{align}
where $H(\mf x,\mf x') = \langle\{ \hat{\phi}(\mf x)\hat{\phi}(\mf x')\}\rangle_{\hat{\rho}_\phi}$ is called the Hadamard function. Since the field commutator is proportional to the identity operator of the field, it is clear that the causal propagator is independent of the field's state $\hat{\rho}_\phi$. The causal propagator encodes information that propagates between different events through the field. The Hadamard function on the other hand, contains all of the state-dependence (including the quantum degrees of freedom) of the field since different field states will not necessarily commute with the anti-commutator of the field.

The qubit-field interaction is prescribed by the interaction Hamiltonian density, in the interaction picture, as
\begin{align}\label{eq:hIappendix}
\hat{h}_d(\mf x) = \lambda_d \Lambda_d(\mf x) \hat{\mu}_d(t) \otimes \hat{\phi}(\mf x),
\end{align}
where the $\lambda$ is a coupling constant, $\hat{\mu}_d(t) = e^{i \Omega t}\ket{1_d}\!\!\bra{0_d}+e^{-i \Omega t}\ket{0_d}\!\!\bra{1_d}$ is the time-evolved operator $\hat{\sigma}_x^{(d)}$ acting on qubit $d$, and $\Lambda_d(\mf x)$ are spacetime smearing functions, which define the shape of the interaction in spacetime. The functions $\Lambda_d(\mf x)$ are each strongly supported around the detectors' trajectories $\mf z_d(t)$. The smearing functions $\Lambda_d(\mathsf x)$ also define smeared versions of the field
\begin{align}
\hat{\phi}(\Lambda_d) = \int \mathrm{d}^4\mf x\, \Lambda_d(\mf x) \hat{\phi}(\mf x)\tbb{.}
\end{align}

The UDW system starts in the initial state $\hat{\rho}(t_0) = \hat{\rho}_{12}(t_0)\otimes \hat{\rho}_\phi = \bigotimes_d^2|0\rangle_d \langle 0| \otimes \hat{\rho}_\phi$, where $\hat{\rho}_\phi$ is chosen to be the vacuum state of the field $|0\rangle_\phi\langle 0|$ and $|0\rangle_d$ are the ground states of $\hat{H}_{d}$. Time evolution of this state to a final time $t$ is given by 
\begin{align}
\hat{\rho}(t) = \hat{U}(t, t_0)\hat{\rho}(t_0)\hat{U}^{\dagger}(t, t_0)\,,
\end{align}
with the time-evolution operators defined by
\begin{align} \label{Equation:TimeEvolve}
\hat{U}(t, t_0) = \mathcal{T}_t\exp{\int^{t}_{t_0} \mathrm{d}^4\mf x\, \hat{h}(\mf x)}\,,
\end{align}
where $\hat{h}(\mf x) = \hat{h}_1(\mf x)+\hat{h}_2(\mf x)$ is the Hamiltonian density of the full system and $\mathcal T_t\exp$ denotes the time-ordered exponential with respect to the time parameter $t$~\cite{Martin-Martinez_Perche_Torres_2021}. 

The spacetime-smearing functions $\Lambda_d(\mf x)$ we consider in this work are given by
\begin{align}
\Lambda_d(\mf x) &= \frac{1}{(2\pi \sigma_d^2)^{3/2} }e^{-\frac{(\bm x-\bar{\bm x}_{\tbb{d}})^2}{2\sigma_d^2}-\frac{(t-\bar{t}_d)^2}{2T_d^2}}\,,
\end{align}
where $T_d \in \mathbb R_+$ is the temporal width for system $d$, $\sigma_d \in \mathbb R_+$ is the spatial width for system $d$, $\bar{t}_d \in \mathbb R$ is the temporal centre for system $d$, and $\bar{\bm x}_{d} \in \mathbb R^3$ is the spatial centre vector for system $d$.

Field-mediated detector correlations have been studied extensively in entanglement harvesting and relativistic quantum communication, where it has been shown that localized detectors can extract correlations from the vacuum and exchange quantum information through a relativistic quantum field~\cite{Valentini1991,Reznik2003,Pozas-Kerstjens_Martin-Martinez_2015,Pozas-Kerstjens_Martin-Martinez_2016,Jonsson_Ried_Martin-Martinez_Kempf_2018,Simidzija_Ahmadzadegan_Kempf_Martin-Martinez_2020}. This literature separates symmetric and causal field correlations and task-dependent information measures computed from detector states. The environment that the RL agent learns in is an RQI setup of two qubit detectors interacting with a real massless scalar quantum field, with the feedback the RL agent receives in the form of a reward function that depends on the quantum mutual information:
\begin{align}
I(i:j)=S(\hat{\rho}_i)+S(\hat{\rho}_j)-S(\hat{\rho}_{ij}),
\label{eq:mutual-info}
\end{align}
where $S(\hat{\rho_i})$ is the von Neumann entropy for the reduced state of system $i$, $\hat{\rho}_i$:
\begin{align}
S(\hat{\rho}_i)=-\Tr\left(\hat{\rho}_i\log(\hat{\rho}_i)\right)\,.
\end{align}
Notice that both $I(i:j)$ and $S(\hat{\rho}_i)$ can be computed explicitly in terms of the spectrum of the operators $\hat{\rho}_i$, $\hat{\rho}_j$ and $\hat{\rho}_{ij}$.

In our protocol, we will consider the case where the detectors have trivial internal dynamics, with $\Omega = 0$. This case is particularly convenient for studying protocols that involve non-infinitesimal coupling constants, as the model can be solved non-perturbatively in these cases~\cite{Landulfo}. Indeed, in Ref. ~\cite{Perche_2024}, the final state of two gapless detectors was found in closed form in terms of smeared Hadamard and retarded Green's functions of the scalar field $\hat{\phi}(\mf x)$ along the detectors' spacetime smearing functions $\Lambda_d(\mf x)$. In particular, if the detector interaction regions are related by a spacetime translation ($\Lambda_1(\mf x) = \Lambda_2(\mf x - \mf v)$) and considering the detectors to be initialized in their ground state ($\hat{\rho}_d(t_0) = \ket{0_d}\!\!\bra{0_d}$), their final state at a time $t$ after the interactions $\hat{\rho}_{12}(t) = \Tr_\phi(\hat{\rho}(t))$ in the $\hat{\sigma}_x$ basis, is given by


\begin{align}
\hat{\rho}_{12}(t) =\frac14
\begin{pmatrix}
1 & e^{-H}e^{-2\ii G_{21}} & e^{-H}e^{-2\ii G_{12}} & e^{-2(H+H_{12})}\\
\ast & 1 & e^{-2(H-H_{12})} & e^{-H}e^{2\ii G_{12}}\\
\ast & \ast & 1 & e^{-H}e^{2\ii G_{21}}\\
\ast & \ast & \ast & 1
\end{pmatrix}
\label{eq:rhoab}
\end{align}
where we defined
\begin{align}
\begin{gathered}
H = \lambda^2H(\Lambda_1,\Lambda_1) = \lambda^2H(\Lambda_2,\Lambda_2),\\
H_{12} = \lambda^2H(\Lambda_1,\Lambda_2)= \lambda^2H(\Lambda_2,\Lambda_1),\\
G_{ij} = \lambda^2G_R(\Lambda_i,\Lambda_j),
\end{gathered}
\end{align} 
with 
\begin{equation}
H(\Lambda_i,\Lambda_j) = \int \dd^4 \mf x \dd^4 \mf x' H(\mf x, \mf x') \Lambda_i(\mf x)\Lambda_j(\mf x'),
\end{equation}
and similarly for $G_R(\Lambda_i,\Lambda_j)$.

Notice that in Eq.~\eqref{eq:rhoab}, the retarded propagators $G_{ij}$ yield non-diagonal phases, while the $H$ and $H_{ij}$ terms are associated with decoherence. Specifically, the $H$ terms determine the local noise experienced by each detector, while the $H_{ij}$ terms correspond to correlated decoherence terms between them. The corresponding reduced states are
\begin{align}
\hat{\rho}_1(t) &= \frac12
\begin{pmatrix}
1 & e^{-H}\cos(2G_{12})\\
e^{-H}\cos(2G_{12}) & 1
\end{pmatrix},\\
\hat{\rho}_2(t) &= \frac12
\begin{pmatrix}
1 & e^{-H}\cos(2G_{21}) \\
e^{-H}\cos(2G_{21}) & 1
\end{pmatrix}.
\end{align}
In particular, the eigenvalues of $\hat{\rho}_{12}$ are
\begin{align}
\lambda_{1,2} &=\frac14-\frac14e^{-2H}\cosh(2H_{12})\\
&\qquad\mp \frac14e^{-2H}\sqrt{4e^{2H}\sin^2(E_{12})+\sinh^2(2H_{12})}\nonumber,\\
\lambda_{3,4} &= \frac14+\frac14e^{-2H}\cosh(2H_{12})\\
&\qquad\mp\frac14e^{-2H}\sqrt{4e^{2H}\cos^2(E_{12})+\sinh^2(2H_{12})}\nonumber\,,
\end{align}
with $E_{ij} = G_{ij} - G_{ji}$ and the reduced eigenvalues are
\begin{align}
\operatorname{Spec}(\hat{\rho}_1(t)) &= \left\{\frac12\left(1\pm e^{-H}\cos(2G_{12})\right)\right\},\\
\operatorname{Spec}(\hat{\rho}_2(t)) &= \left\{\frac12\left(1\pm e^{-H}\cos(2G_{21})\right)\right\}.
\end{align}
We can see that the quantum mutual information only depends on the following parameters: the coupling constant $\lambda $, the spatial smearing width $\sigma_1 = \sigma_2 = \sigma$, the temporal smearing width $T_1 = T_2 = T$, the spatial smearing centre $\bar{\bm x}_2$, and the temporal smearing centre $\bar{t}_2$. In this work we will fix $\lambda, \sigma, T$ while allowing the relative spacetime position of the detectors (encoded in $\bar{t}_2$ and $\bar{\bm x}_2$) to be varied. 

The core idea of our paper is to allow an RL agent to move the second detector relative to the first, attempting to maximize the quantum mutual information between the two detectors after their interaction with the field. More precisely, we will describe $\bar{t}_2$ and $\bar{\bm x}_2$ as values in a spacetime grid, constituting the RL agents' environment. The RL agent then explores the environment by changing the spacetime placement of the second detector, receiving a reward function based on how well it is able to improve the quantum mutual information through local changes in $\bar{t}_2$ and $\bar{\bm x}_2$. This task updates the RL agent's model of the environment and updates its memory at each step. We will discuss the task structure and the RL agent's model in further detail in the following two sections.



\section{Reinforcement-Learning and Agent Architecture} \label{Section:RL}

\begin{figure*}[t!]
    \centering
    \includegraphics[width=\textwidth]{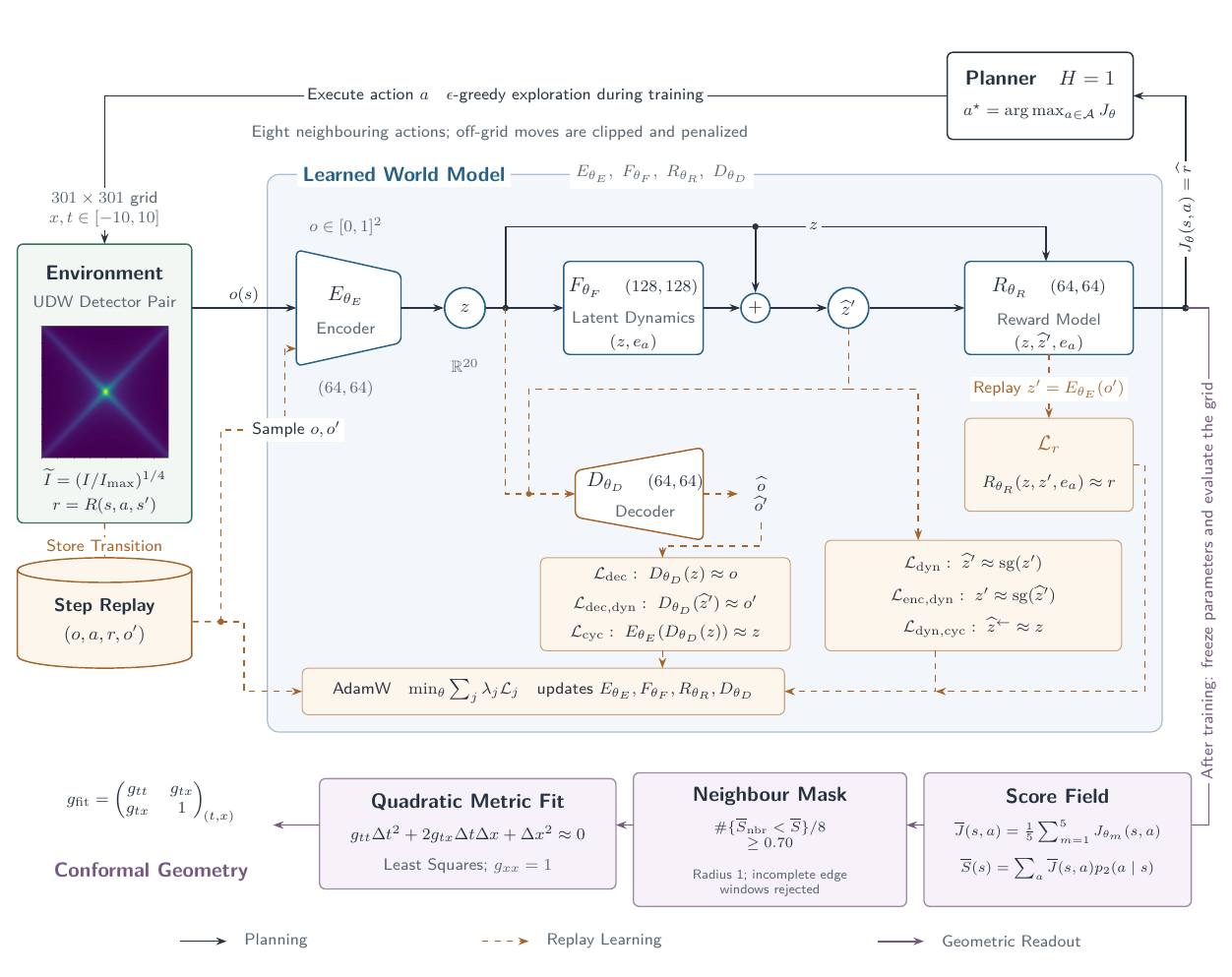}
    \caption{A schematic of the agent architecture and learning pipeline. Detector 1 is fixed at the origin and the $(x,t)$ grid represents the spacetime smearing centres of detector 2. The thumbnail is Fig.~\ref{fig:landscape}, showing the transformed quantum mutual information contained in the reward (Eq.~\eqref{eq:reward-active}) received by the agent. All eight local actions remain available during planning and exploration, with proposed off-grid moves being clipped by the environment and given a boundary penalty. Here $e_{a_t}$ is the one-hot representation of the action, and numbers beside network symbols specify hidden-layer widths, while dashed connections show step-replay learning with shared network parameters. The RL agent chooses these actions by initially using the encoder to map the normalized observation $o_t$ to the latent state $z_t=E_{\theta_E}(o_t)\in\mathbb R^{20}$, then the residual dynamics predicts $\widehat z_{t+1}=z_t+F_{\theta_F}(z_t,e_{a_t})$ which is then fed into the reward model that predicts $\widehat r_{t+1}=R_{\theta_R}(z_t,\widehat z_{t+1},e_{a_t})$. Since the planning horizon is $H=1$, the planner score is simply $J_\theta(s_t,a_t)=\widehat r_{t+1}$. Reward training uses $z_{t+1}=E_{\theta_E}(o_{t+1})$, the encoded observed next state, whereas planning uses the model-predicted state $\widehat z_{t+1}$. The decoder acts separately on encoded and predicted latent states. Its two displayed outputs are the reconstructed observation $\widetilde o_t=D_{\theta_D}(z_t)$ and predicted next observation $\widehat o_{t+1}=D_{\theta_D}(\widehat z_{t+1})$;
coordinate reconstruction is also applied to the encoded next observation $z_{t+1}$. The seven loss constraints Eqs.~ \eqref{eq:reward-loss},\eqref{eq:dynamics-loss}, and ~\eqref{eq:dynamics-consistency-losses}- \eqref{eq:decoder-losses} are shown alongside their corresponding predictions, with $\operatorname{sg}$ denoting stop-gradient and $\widehat z^{\leftarrow}_t$ the reverse-action prediction obtained from $\widehat z_{t+1}$. The $\lambda_j$ denote the corresponding loss weights. The purple branch first averages the planner action-score fields from the five trained agents into $\overline J(s,a)$ and then forms the scalar softmax-average field $\overline S$. A cell enters the candidate lightcone mask when its score exceeds that of at least $70\%$ of the eight neighbouring cells in a radius-1 square. Cells without a complete neighbourhood are rejected. Least squares then fits the quadratic form to the extracted branch points, with displacements measured relative to the fixed detector and the conformal freedom fixed by $g_{xx}=1$, yielding a representative of the conformal metric in $(t,x)$ coordinate order.}
    \label{fig:agent-architecture}
\end{figure*}

In this section we will introduce the reinforcement-learning (RL) framework and RL agent architecture needed to formulate the learning task and to analyze the RL agent’s learned representation in the next sections. We first introduce some basics of RL that are required for the present work before describing the model architecture used by the RL agent, which is shown schematically in Fig.~\ref{fig:agent-architecture}. Readers already familiar with RL may proceed directly to the description of the model architecture below, and those already versed in machine learning more generally are welcome to proceed directly to the next section without loss of understanding. 

As we are interested in an agentic perspective on learning spacetime structure, the problem is naturally formulated in terms of RL. The basic agent–environment interaction loop of RL starts with the agent observing a state and deciding to select an action, then receiving a reward as feedback, and finally updating a memory function via a learning update rule whose purpose is to maximize future rewards ~\cite{sutton2018reinforcement}. In the present setting the state is a spacetime smearing centre for detector 2, the action is a local displacement on the spacetime grid, and the reward is a function of the quantum mutual information, so the RL formulation keeps the operational character of the task explicit.

The RL task is formalized as a Markov Decision Process (MDP), written as the 4-tuple $(\mathcal S,\mathcal A,\mathcal R, P)$. Here $\mathcal S$, $\mathcal A$, $\mathcal R$ are the state, action, and reward spaces, respectively; we will generally denote their elements as $s\in \mathcal S$, $a \in \mathcal A$, $r \in \mathcal R$. The full transition kernel $P$ specifies the probability of each possible next state and reward, given the current state and the chosen action:
\begin{align}
P(s',r\mid s,a)=\mathbb{P}(S_{t+1}=s',R_{t+1}=r\mid S_t=s,A_t=a)\,,
\end{align}
with $S_t, A_t, R_t$ denoting random variables at time $t$ who can take values from the state, action, and reward spaces, respectively. From these quantities we can also define the state-transition kernel (where we keep the symbol $P$, but omit its second argument) as
\begin{align}
P(s'\mid s,a)=\sum_{r \in \mathcal R} P(s',r\mid s,a)
\end{align}
and the expected reward function from a reward being delivered when transitioning from state $s$ to $s'$, $s\rightarrow s'$ while taking the action $a$, as
\begin{align}
R(s,a,s') &= \mathbb E[R_t\mid S_t = s, A_t = a, S_{t+1}=s' ] \nonumber\\
&= \sum_{r \in \mathcal R} r \frac{P(s',r\mid s,a)}{P(s'\mid s, a)}\,,
\end{align}
where $\mathbb E[\cdot|\cdot]$ is the conditional expectation value. The behaviour of the RL agent (i.e. how it will act given some state) at some time $t$ can be described by the stochastic policy $\pi(a\mid s)=\mathbb{P}_\pi(A_t=a\mid S_t=s)$, and the RL agent's goal is typically to maximize their expected discounted return (called the value function)
\begin{align}
v_\pi(s) = \mathbb E_\pi[G_t\mid S_t = s]
\end{align}
where
\begin{equation}
G_t=\sum_{k=0}^{\infty}\gamma^kR_{t+k+1}
\label{eq:discounted-return}
\end{equation}
is called the discounted return and $\gamma\in[0,1)$ is called a discount factor, which controls the strength of future rewards~\cite{bellman1957dynamic,sutton2018reinforcement}; $\gamma$ both makes the return finite under standard bounded-reward assumptions and determines how strongly later detector placements influence present actions.

There are two broad ways in which an RL agent can use experience to improve its decisions: model-free value-based methods and model-based methods. The model-free value-based methods, such as Q-learning~\cite{watkins1989learning,watkins1992qlearning}, directly learn the expected long-term return associated with taking an action in a state, which works quite well for small- to moderate-sized problems. For large state spaces, these value functions can be more efficiently represented by artificial neural networks, which provide flexible parameterized function approximators capable of generalizing between states rather than storing a separate value for every state-action pair~\cite{cybenko1989approximation,hornik1989multilayer,mnih2013playing,mnih2015human}. Model-based methods on the other hand, have the RL agent instead learn an internal predictive model of the environment and use that model to evaluate possible actions before executing them~\cite{sutton2018reinforcement}. 

We adopt the model-based approach to model the RL agent's memory and decision-making faculties, which in our case, consists of an encoder, decoder, transition model, reward model, and a planning rule that combines their predictions into a decision (see Fig. \ref{fig:agent-architecture} for how these components fit together). We will refer to the collection of these learned internal models as the \textit{agent's memory}, since it is the orchestration of these learned models that retains the necessary behavioural information to act in the environment. This model-based approach is important for our purposes because the learned agent model provides explicit internal objects that can be interrogated after training, meaning that we can also examine whether the representations and predictions supporting its behaviour retain information about the spacetime structure that generated the reward landscape.

To distinguish the physical state from its representations within the agent, let $s_t\in\mathcal S$ denote the physical environment state and $o_t=o(s_t)\in\mathcal O$ the normalized observation supplied to the encoder neural network (this will be shown in the next section) at step $t$. In the present problem, $s_t$ specifies the spacetime smearing centre of detector 2 and $o_t$ is a normalized version of it. The decision-making process begins with the encoder, which maps this observation to an internal latent state
\begin{align}
z_t = E_{\theta_E}(o_t)\in\mathcal Z,\qquad\mathcal Z\subseteq\mathbb R^{d_z},
\label{eq:latent-encoder}
\end{align}
where $\theta_E$ denotes the trainable parameters of the encoder, $d_z$ is the latent dimension, and the latent space $\mathcal Z$ is internal to the agent, having elements that are internal vector representations of the observations whose individual components are not assigned any intrinsic physical or geometric meaning. It is the encoder that determines which information about the observation is retained in the internal representation and must preserve enough information to support later dynamics and reward prediction, as well as reconstruction by the decoder.

The RL agent's decision-making then proceeds through planning over this latent representation. A general finite-horizon model-based planner considers candidate action sequences $a_{0:H-1}$ comprised of $H$ simulated future actions ($H$ is called the planning horizon), by internally predicting the transitions and rewards generated by each sequence, and subsequently executes the first action of the sequence with the largest predicted discounted return. Here and below, the index $t$ labels actual interactions with the environment, whereas $k=0,\ldots,H$ labels model-predicted steps within one of these planning rollouts (defined further down), and we use hats to denote quantities predicted by the learned model. Starting from the encoded current state $\widehat z_0=z_t$, the planner generates the model-predicted latent trajectory
\begin{align}
\widehat z_{k+1}=\widehat z_k+F_{\theta_F}\!\left(\widehat z_k,e_{a_k}\right),
\label{eq:latent-dynamics}
\end{align}
where $F_{\theta_F}$ is the learned latent dynamics model with trainable parameters $\theta_F$ and the vector $e_{a_k}$ is a one-hot encoding of action $a_k$ (meaning that its component corresponding to $a_k$ is one and all remaining components are zero). We choose the form of Eq.~\eqref{eq:latent-dynamics} because it makes $F_{\theta_F}$ learn action-dependent displacements in latent space rather than just predicting the next latent state. 

For the initial planning step $\widehat o_0=o_t$, while subsequent predicted observations are obtained from the decoder $D_{\theta_D}:\mathcal Z\rightarrow\mathcal O$, which maps latent states back to observed normalized coordinates and has trainable parameters $\theta_D$. To distinguish reconstructions of actual observations from model-predicted future observations, we use tildes to denote the reconstruction of the current observation,
\begin{align}
\tilde o_t = D_{\theta_D}(z_t)
\label{eq:decoder}
\end{align}
whereas a decoded model prediction during planning is denoted with a hat
\begin{align}
\widehat o_{k+1} = D_{\theta_D}(\widehat z_{k+1})\tbb{.}
\label{eq:predicted-decoder}
\end{align}
The role of the decoder encourages
\begin{align}
D_{\theta_D}(E_{\theta_E}(o_t))\simeq o_t\,,
\end{align}
i.e. it encourages that decoded predicted next states should lie near the corresponding observed next coordinates. The decoder anchors the latent representation to the detector coordinates without imposing a particular geometry upon that representation, which is a key distinction because although the encoder receives detector coordinates as input, the latent variables themselves are not amenable to a geometric interpretation \emph{a priori}.

The last learned component is the reward model $R_{\theta_R}$, which during training predicts the next reward given the current latent state $z_k$ and the encoded observed next latent state $z_{k+1} = E_{\theta_E}(o_{k+1})$ under action $a_k$, as
\begin{align}
r'_{k+1} = R_{\theta_R}\!\left(z_k, z_{k+1},e_{a_k}\right),
\label{eq:local-reward-model-training}
\end{align}
with $\theta_R$ denoting its trainable parameters. The planner instead uses the model-predicted transition $\widehat z_k\rightarrow\widehat z_{k+1}$ under action $a_k$ since the future state $z_{k+1}$ is unavailable to them, which is given by
\begin{align}
\widehat r_{k+1} = R_{\theta_R}\!\left(\widehat z_k,\widehat z_{k+1},e_{a_k}\right).
\label{eq:local-reward-model-planning}
\end{align}
This procedure thus channels all of the previous machinery into building the representation $\widehat r$ that is finally used to assign a candidate action sequence the score
\begin{align}
J_\theta(s_t,a_{0:H-1})=\sum_{k=0}^{H-1}\gamma^k\widehat r_{k+1}, \qquad\theta=(\theta_E,\theta_F,\theta_D,\theta_R),
\label{eq:planner-score}
\end{align}
and then promptly select the action\footnote{The dependence of the score on the other trainable parameters besides $\theta_R$ comes in implicitly through the latent vector representations $\widehat z_k$ that $\widehat r_k$ is a function of.}
\begin{align}
a_t^\star=\left[\arg\max_{a_{0:H-1}}J_\theta(s_t,a_{0:H-1})\right]_0.
\label{eq:planner-action}
\end{align}
Importantly, a model-predicted rollout here refers to a sequence generated internally by the learned dynamics and reward models rather than one actually executed in the environment. In the present work we use the planning horizon $H=1$, meaning that the planner predicts the next latent state and immediate reward associated with each valid local action from the current encoded state $z_t$, then choosing the action with the largest predicted reward
\begin{align}
a_t^\star=\arg\max_{a\in\mathcal A}R_{\theta_R}\!\left(z_t, z_t+F_{\theta_F}(z_t,e_a),e_a\right).
\label{eq:horizon-one-planner}
\end{align}
Thus, although the observations supplied to the encoder are detector coordinates, action evaluation is mediated entirely by the learned latent representation and predictive models.


To learn the consequences of a broad range of actions, the RL agent must sometimes explore alternatives to the planner’s recommendations. The agent therefore follows the planner’s recommendation with probability $1-\epsilon$ and selects an action uniformly at random with probability $\epsilon$, where $0\leq\epsilon\leq1$. For a specified constant $\epsilon$, this is implemented by drawing a number $u$ uniformly from $[0,1]$: if $u\geq\epsilon$, the agent follows the planner’s recommendation; otherwise, it samples an action uniformly from the available actions. This procedure is known as $\epsilon$-greedy exploration. An exploration schedule is also implemented, in which $\epsilon$ varies with the cumulative number $n$ of environment steps according to $\epsilon(n)=0.1+0.9e^{-n/\epsilon_{\text{decay}}}$. This schedule starts at $\epsilon=1$ and gradually approaches $0.1$, so that early training is dominated by uniformly random actions that populate the replay memory with a broad range of local transitions (more on this below), while later training increasingly uses the planner’s output.

Figure \ref{fig:agent-architecture} demonstrates how the learning process unfolds through each of the above components. The training and associated loss functions will be described in the next section.

\section{Learning Task} \label{Section:LearningTask}

In this section we introduce and describe the details of the learning task and parameter choices of the RL agent architecture. The question we are interested in addressing in this work is whether a learning agent performing a quantum information task that is embedded in spacetime --- with no \textit{a priori} notion of spacetime --- can develop a representation of that spacetime within its memory after successfully completing the task. More concretely, given the predictive models and behaviour policy of a (model-based) RL agent tasked with maximizing a monotonic function of the quantum mutual information between two UDW qubit systems, both embedded in a spacetime, can we extract the causal or metric structure of the spacetime from them? The simplicity of the UDW model, combined with how it connects quantum information and spacetime geometry, makes it an ideal choice for setting up the problem environment. Note that from now on, we will hold two of the spatial dimensions of the $(3+1)$ dimensional spacetime fixed and consider a $(1+1)$-dimensional slice.

In our setting, the MDP that represents the task, $(\mathcal S,\mathcal A,\mathcal R,P)$, has a state space $\mathcal S$ consisting of a $301\times 301$ grid whose cells index spacetime smearing centres
\begin{align}
\mf x=(x,t),\qquad x,t\in[-10,10],
\end{align}
for detector $2$, with detector $1$ fixed at the origin. If $i_x,i_t\in\{0,\ldots,300\}$, the physical coordinates are
\begin{align}
x(i_x) &=-10+i_x\Delta,\\
t(i_t) &=-10+i_t\Delta,\\
\Delta &=\frac{20}{301-1},
\label{eq:grid-coordinates}
\end{align}
and the observed input to the model at interaction step $t$ is the normalized coordinate representation
\begin{align}
o_t=o(s_t)=\left(\frac{x+10}{20},\frac{t+10}{20}\right)\in[0,1]^2.
\label{eq:coordinate-normalization}
\end{align}
 The goal state of the task corresponds to the centre of the physical grid $(0, 0)\in[-10,10]^2$ where the detectors' interaction regions coincide. Since quantum mutual information measures the total information content of the joint system excluding correlations between subsystems, it is maximized in our setting when the detectors' interactions are as close as possible; the reason being that field-mediated correlations and communication can decay rapidly with detector separation. An episode terminates when the RL agent reaches this goal state, corresponding to the maximum transformed quantum mutual information $\tilde I_{\max}$, or when the maximum number of allowed steps is reached. An episode is considered successful if the RL agent reaches the goal state before this maximum.

The action set $\mathcal A$ consists of the eight nearest-neighbour grid moves
\begin{align}
\mathcal A=\{(\delta_x,\delta_t):\delta_x,\delta_t\in\{-1,0,1\}\}\setminus\{(0,0)\},
\label{eq:action-set}
\end{align}
so that an action changes the integer cell by $(i_x+\delta_x,i_t+\delta_t)$, corresponding to a physical displacement of $(\Delta x,\Delta t)=(x(i_x+\delta_x)-x(i_x),t(i_t+\delta_t)-t(i_t))$. Notice that each state in $\mathcal{S}$ corresponds to a run of the experiment where the separation between the detectors is $(x,t)$. The agent's decision to move the relative interaction centres by $(\Delta x, \Delta t)$ then corresponds to another run of the experiment associated with the new relative separations $(x + \Delta x, t + \Delta t)$. The environment also prevents proposed moves that would leave the grid and assigns a corresponding boundary penalty $b_t = 0.2\,\mathbf 1_{\rm boundary}(s_t,a_t)$. The spacetime grid transition itself is deterministic, and the stochasticity during training comes from the $\epsilon$-greedy exploration rule rather than from stochastic spacetime dynamics or from sampling the detector state.

\begin{figure}[t!]
    \centering
    \includegraphics[width=0.49\textwidth]{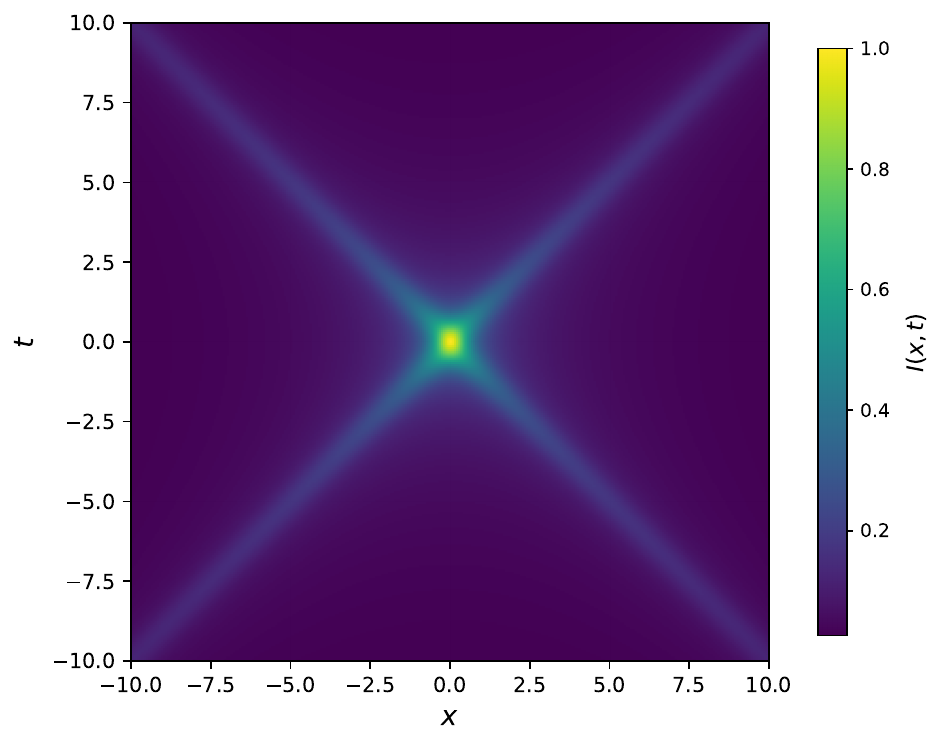}    \caption{
    The transformed quantum mutual information landscape Eq.~\eqref{eq:mi-transform}, navigated by the agent. The plot shows the broad central high-information region corresponding to small detector separation and the diagonal ridge structure associated with the light cone that the RL agent samples only through the reward.
    }
    \label{fig:landscape}
\end{figure}

To increase the contrast and visibility of the quantum mutual information further away from the centre of the spacetime grid without breaking its monotonicity as an information measure, we use the normalized power transform of the quantum mutual information
\begin{align}
\tilde I(\mf x) = \left[\frac{I(\mf x)}{\max_{\mf y} I(\mf y)}\right]^p,\qquad p=\frac{1}{4}.
\label{eq:mi-transform}
\end{align}
The detector parameters used for this landscape are\footnote{In principle $T$ and $\sigma$ must have units of length and time. Throughout the manuscript we write them assuming an arbitrary timescale $T_0$. For instance, we have $T = 0.15T_0$ and $\sigma = 0.15 c T_0$ in this example, where $c$ is the speed of light. Our results are valid for any choice of $T_0$ due to the conformal character of the massless field.} $T=0.15$, $\sigma=0.15$, and $\lambda=1$, with the initially uncorrelated product detector state given in Sec.~\ref{Section:UdW}. The bare quantum mutual information range is $[3.1929\times10^{-8}, 6.5477\times10^{-2}]$ and the transformed range is approximately $[0.02643,1]$. Figure~\ref{fig:landscape} visualizes this function, highlighting the dependence of the quantum mutual information on the properties of spacetime through the central, diagonal lightcone ridges.

The reward from the environment is comprised of four terms: 
\begin{align}
R(s_t,a_t,s_{t+1}) &= \big[\tilde I(s_{t+1})-\tilde I_{\max}\big] + 0.02\,\mathbf 1_{\Delta\tilde I_t\ge0}\nonumber\\
&\quad - 0.05\,\mathbf 1_{\Delta\tilde I_t<0} - b_t,
\label{eq:reward-active}
\end{align}
where the realized reward is $r_{t+1}=R(s_t,a_t,s_{t+1})$, $b_t$ is the boundary penalty defined previously, $\Delta \tilde I_t = \tilde I(s_{t+1})-\tilde I(s_t)$, $\tilde{I}_{\text{max}} = 1$ in this case, and no additional terminal reward is assigned upon reaching the goal state. The first term merely rewards the RL agent for being in large magnitude states and subtracting off the maximum from it essentially creates a penalty field, while termination upon reaching the maximum prevents the RL agent from exploiting the reward by hovering near the goal state. The next two terms, $0.02\,\mathbf 1_{\Delta\tilde I_t\geq0}$ and $-0.05\,\mathbf 1_{\Delta\tilde I_t<0}$, measure the progress of the RL agent towards the goal state, with the former term giving a small positive reward for moves that do not decrease the quantum mutual information and the latter term giving a slightly larger penalty for moves that do. The reward inherits the dependence on spacetime structure from the quantum mutual information, and since the RL agent does not know what this function is or represents, from the agent’s perspective, this feedback is only implicitly associated with said structure.

\begin{figure*}[t!]
    \centering
    \includegraphics[width=0.9\textwidth, trim=5 5 10 10, clip]{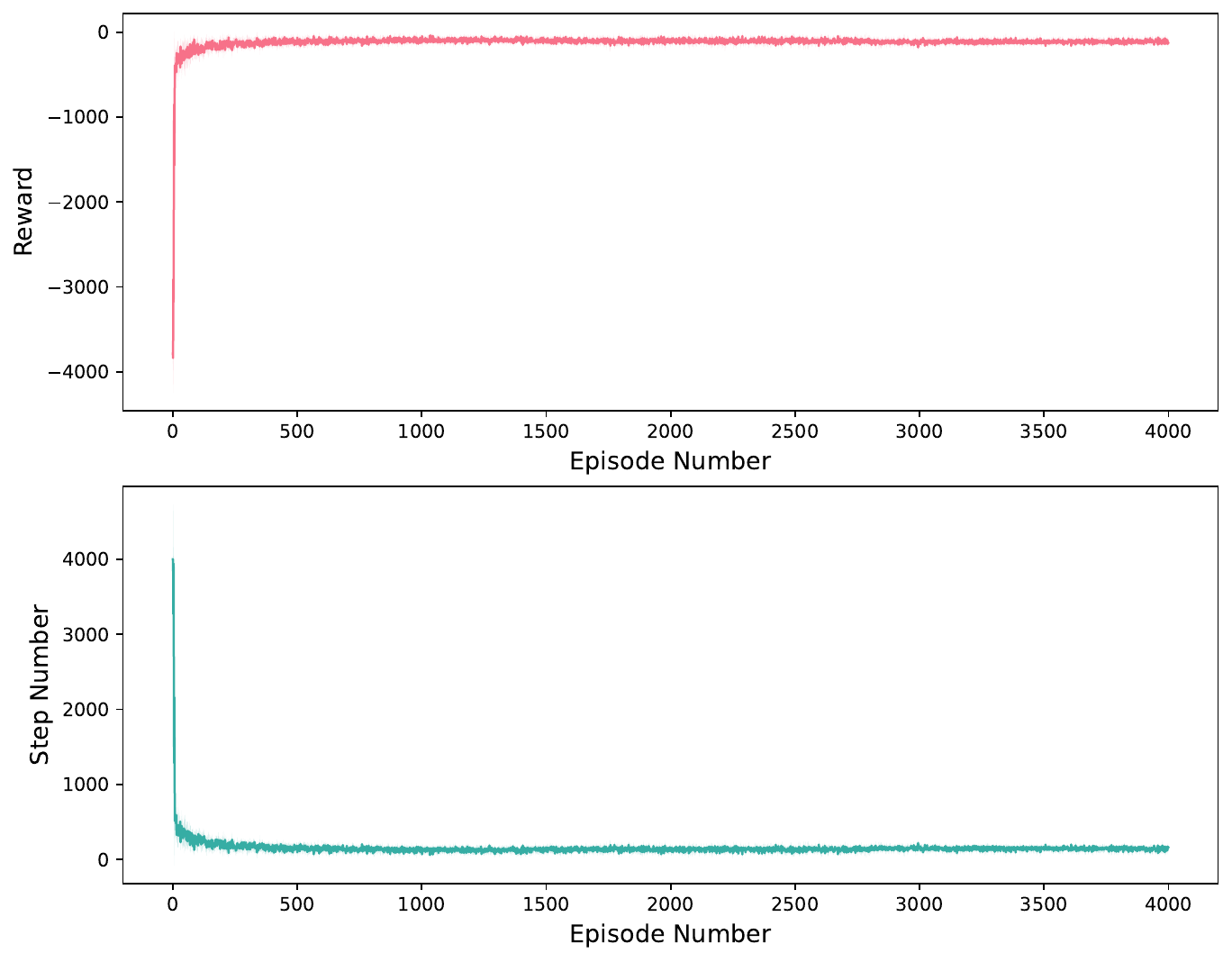}
    \caption{
    Learning curves for the model. On top is the episode return and below that is the episode length over $4000$ episodes, both showing rapid convergence to the optimal policy and stability thereafter. Each plot is an average over 5 agents with the shaded regions representing the standard deviation around the average curve with respect to the 5 agents.
    }
    \label{fig:training}
\end{figure*}

The purpose of using a model-based architecture in this work is to add an extra element of genuine discovery of the environment structure by the agent. The reward already masks the explicit structure, but the RL agent still receives spacetime coordinates, and so the job of the model is to first translate them into an internal latent state via the trainable encoder (only implicitly dependent on spacetime coordinates), whereas the job of the decoder is to make this representation somewhat faithful. After this point, the RL agent does all of its decision-making based on the internal latent states: training a reward model of the environment reward and a latent dynamics model of the environment transition probability to simulate the environment from their own internal viewpoint. In this sense, the model really can be argued to be discovering the spacetime structure autonomously from scratch.


During training, the RL agent learns to navigate the environment by updating its models from feedback via the minimization of a composite \textit{loss function} over uniformly sampled experience replay mini-batches. Experience replay works as a memory buffer that holds observed transitions $(s_t,a_t,r_{t+1},s_{t+1})$----including episode-boundary markers so that sampled windows do not cross resets----from which randomly sampled subsets of one-step transitions or mini-batches are subsequently used for parameter updates~\cite{lin1992self,mnih2015human}. A sampled mini-batch contains transitions $(o_{b,k},a_{b,k},r_{b,k+1},o_{b,k+1})$ consisting of the current normalized coordinates $o_{b,k}$, action $a_{b,k}$, realized reward $r_{b,k+1}$, and next normalized coordinates $o_{b,k+1}$, with batch index $b$. We denote encoded latent states $z_{b,k}$ and predicted next latent states $\widehat z_{b,k+1}$ by
\begin{align}
z_{b,k}&=E_{\theta_E}(o_{b,k}),\\
z_{b,k+1}&=E_{\theta_E}(o_{b,k+1}),\\
\widehat z_{b,k+1} &= z_{b,k}+F_{\theta_F}\left(z_{b,k},e_{a_{b,k}}\right),
\end{align}
and use $\langle\cdot\rangle$ for the uniform average over the sampled replay mini-batch. The norm we will use for the loss functions below is the mean-square norm $\|v\|^2=d^{-1}\sum_{i=1}^d v_i^2$ for vectors $v\in\mathbb R^d$. The composite loss function is a linear combination of a series of individual loss functions that play different roles. Let us now go through each term of the composite loss function.

The first loss term is the reward-prediction loss, and it trains the local reward model $r'_{b,k+1}$ on the immediate realized reward $r_{b,k+1}$:
\begin{align}
\mathcal L_r= \left\langle\left(r'_{b,k+1}-r_{b,k+1}\right)^2\right\rangle,
\label{eq:reward-loss}
\end{align}
where
\begin{align}
r'_{b,k+1} = R_{\theta_R}\!\left(z_{b,k}, z_{b,k+1},e_{a_{b,k}}\right)
\end{align}
is computed from $z_{b,k}$, $ z_{b,k+1}$, and $a_{b,k}$. The next loss term is the latent dynamics loss, and it trains the latent transition model to land near the encoded next observation:
\begin{align}
\mathcal L_{\rm dyn}=\left\langle\left\|\widehat z_{b,k+1}-\operatorname{sg}(z_{b,k+1})\right\|^2\right\rangle,
\label{eq:dynamics-loss}
\end{align}
where $\operatorname{sg}(\cdot)$ denotes the stop-gradient operation, which blocks the flow of gradients during optimization so that the encoded target is detached in the main dynamics term and the transition model does not chase a moving encoder. The next two loss terms act as consistency constraints, keeping this separation from becoming a loophole. If $a^{-1}_{b,k}$ denotes the reverse grid action, then the corresponding latent dynamics update is 
\begin{align}
\widehat z^{\leftarrow}_{b,k} &= \widehat z_{b,k+1} + F_{\theta_F}\!\left(\widehat z_{b,k+1},e_{a^{-1}_{b,k}}\right),
\end{align}
and the loss terms are given by
\begin{align}
\mathcal L_{\rm dyn,cyc} &= \left\langle\left\|\widehat z^{\leftarrow}_{b,k}-z_{b,k}\right\|^2\right\rangle,\\
\mathcal L_{\rm enc,dyn} &=\left\langle\left\|\operatorname{sg}(\widehat z_{b,k+1})-z_{b,k+1}\right\|^2\right\rangle .
\label{eq:dynamics-consistency-losses}
\end{align}
The first term enforces an imagined forward step followed by the reverse action to return to the original latent state, while the second term applies the same transition mismatch with the predicted latent state detached, so that the encoder is also pressured to remain compatible with the latent dynamics model. The last batch of loss terms are associated with the decoder, and their job is to anchor the latent state to the observed coordinates:
\begin{align}
\mathcal L_{\rm dec} &= \left\langle\left\|D_{\theta_D}(z_{b,k})-o_{b,k}\right\|^2\right\rangle,\\
\mathcal L_{\rm dec,dyn} &= \left\langle\left\|D_{\theta_D}(\widehat z_{b,k+1})-o_{b,k+1}\right\|^2\right\rangle,\\
\mathcal L_{\rm cyc} &= \left\langle\left\|E_{\theta_E}(D_{\theta_D}(z_{b,k}))-z_{b,k}\right\|^2\right\rangle.
\label{eq:decoder-losses}
\end{align}
The first term reconstructs the normalized coordinate features from encoded latent states, the second term tries to constrain predicted next latent states to decode near the observed next coordinate, and the third term prevents the decoder from being accurate only after leaving the encoder's latent representation. Putting all these loss terms together, the composite loss function is therefore
\begin{align}
\mathcal L_{\rm total} &= \mathcal L_r + 2.5\,\mathcal L_{\rm dyn} \nonumber\\
&\quad + 0.2\,\mathcal L_{\rm dec} + 0.2\,\mathcal L_{\rm dec,dyn}+ 0.05\,\mathcal L_{\rm cyc} \nonumber\\
&\quad + 0.5\,\mathcal L_{\rm dyn,cyc} + 0.8\,\mathcal L_{\rm enc,dyn}.
\label{eq:latent-planner-loss}
\end{align}
As is standard practice when working with deep neural networks, the weights attached to the loss function terms are tuned so the learning scheme can be successful. The largest weight falls on the dynamics term because the planner needs a usable one-step latent model before its score surface can carry meaningful action information. The reward loss ties the latent transition to the environment reward, which is crucial for properly navigating the environment and completing the task, so it has a weight of 1. The latent transition loss terms also have relatively large weights because of the previously stated importance of the transition model. Finally, the decoder loss terms keep the latent representation tied to the observed detector grid, which is useful but not as important as the other terms, so they have small weights.

The training hyperparameters we used are as follows. The run uses $N_{\rm ep}=4000$ episodes with at most $T_{\max}=4000$ environment steps per episode, which fixes the maximum number of detector placements the RL agent can sample before an episode terminates. Since the planner uses horizon $H=1$, its score contains only the predicted immediate reward, so $\gamma=0.97$ does not affect action selection and the reported episode return is likewise the undiscounted sum $\sum_t r_{t+1}$ of the shaped rewards. As we mentioned earlier, exploration is $\epsilon$-greedy with $\epsilon_{\rm start} = \epsilon(0) =1.0$, $\epsilon_{\rm final} \sim 0.1$ (with a maximum number of steps $n_\text{max}= N_\text{ep} T_{\text{max}}$), and a step-based decay constant $\epsilon_{\rm decay}=8.0\times10^4$. Episode starts are sampled uniformly during the first $35\%$ of training, from a $70\%$ uniform and $30\%$ edge-biased mixture during the next $35\%$, and from a $40\%$ uniform and $60\%$ edge-biased mixture during the final phase. In each phase, the stated probabilities first select the corresponding region and the starting state is then sampled uniformly from that region, with edge starts drawn from a band of $25$ cells near the grid boundary. This start schedule forces the planner to learn from longer moves across the finite grid rather than only from starts near the central high-information region, which assists in grid coverage and reinforcing far away states. The latent dimension is set to $d_z=20$ so the internal latent representation can be rich enough. The stochastic gradient-descent optimization is implemented by the AdamW algorithm from PyTorch with a batch size for experience replay of $B=256$, a learning rate for the AdamW step size of $\eta=10^{-4}$, a weight decay of $\lambda_{\rm wd}=10^{-5}$, and a gradient-norm cutoff of $\|\nabla\|_{\max}=7$ to prevent occasional high-error replay batches from destabilizing the learned model. The stated weight decay is applied to multidimensional weights, while biases and one-dimensional parameters have zero weight decay, otherwise all other AdamW settings retain their PyTorch defaults. The step-replay memory has capacity $C_{\rm replay}=5\times10^4$, and replay updates begin once it contains $N_{\rm replay}^{\min}=4000$ usable transition windows. Each eligible environment step performs one gradient update, and after the first 50 episodes, the update interval is selected adaptively from $(1,4,8,16,32)$ steps using the preceding 50 episodes. The adaptive routine requires a recent success rate of at least $0.9$, a recent mean episode length no more than $1.25$ times the lower quartile among successful episodes, a recent mean return no more than $0.10$ robust return-scale below the historical upper quartile, and a recent mean prediction error no more than $1.25$ times its historical median. The interval is set by the number of satisfied criteria, with a failure returning the schedule to the initial one-step interval. Each of the predictive objects uses a multi-layered neural network with SiLU activation functions
\begin{align}
\text{SiLU}(x) = \frac{x}{1+e^{-x}}
\end{align}
and linear output layers. The encoder maps the two-dimensional observation to the $d_z=20$ dimensional latent state, while the decoder maps the latent state back to the two-dimensional observation. The encoder, decoder, and reward models all have hidden-layer widths of $(64,64)$, and the latent dynamics model uses $(128,128)$. From the definitions above, the dynamics model takes the latent state and one-hot action as input and outputs a $d_z$-dimensional latent displacement, while the reward model takes the current and encoded observed next latent states together with the one-hot action and outputs a single predicted reward. 

Training is performed for five different agent-parameter and environment random-generator seed initializations, and the final diagnostic fields used for plotting and metric recovery are averaged over these trained runs. The RL agent seed controls the network initialization, exploration, replay mini-batch sampling, and planner sampling, and edge/arm curriculum starts, while the environment seed controls selection of the starting state. After each training run, the network parameters are frozen and the model is evaluated over $1000$ episodes with exploration disabled and starting states sampled uniformly from the full grid with a separate seed. The maximum number of allowed steps remains $T_{\max}=4000$, and an evaluation episode is successful when the RL agent reaches the goal state $(0,0)$ before this limit. Every trained RL agent reaches a success rate of $1.0$ and a mean final objective $\tilde I=1.0$. The average of the five evaluation summaries gives a mean episode length of approximately $103.85$ and a mean undiscounted shaped return of approximately $-76.85$. The training loop can be visualized in Fig. \ref{fig:agent-architecture} and the learning curves in Fig. \ref{fig:training}.

\section{Geometric Features of the Agent Memory} \label{Section:AgentMemory}


Figure \ref{fig:training} shows that the RL agent has essentially converged to the optimal policy for completing the task. If the predictive models supporting this converged policy have retained the spacetime dependence encoded indirectly in the reward landscape, this structure should therefore be detectable in its transition $F_{\theta_F}$ and reward $R_{\theta_R}$ model memories as well as its internal latent state. The planner uses these learned objects to predict the outcomes of candidate local actions and optimize the score Eq. \eqref{eq:planner-score}, allowing us to explicitly inspect whether its decisions retain information about the causal structure underlying the detector environment. The question is then: do the relations within these objects retain enough information for the causal geometry underlying the detector environment to be reconstructed? We train five independent agents, indexed by $m=1,\ldots,5$, and average the corresponding diagnostic fields over these agents unless otherwise stated.

We will study the memory of the planner to try and answer the above question using two fields that provide complementary views of this learned memory. For each spacetime grid point $\mathsf{x}=(x,t)$, let $s(\mathsf{x})$ denote its corresponding environment state. The first diagnostic is the best-action planner score
\begin{align}
\overline B(\mathsf{x}) &= \frac{1}{5}\sum_{m=1}^{5}B^{(m)}(\mathsf{x}),
\label{eq:best-planner-score}
\end{align}
which is the average over the best-action planner scores (Eq.~\eqref{eq:planner-score}) for each RL agent $m$:
\begin{align}
B^{(m)}(\mathsf{x}) &=\max_{a\in\mathcal A}J_{\theta^{(m)}}(s(\mathsf{x}),a),
\label{eq:best-planner-score-agent}
\end{align}
where the active settings use the full eight-action set $\mathcal A$ at every $\mathsf{x}$, so this assigns to each detector placement the score of its most favourable action $a^*_t$ (Eq. \eqref{eq:planner-action}). The second diagnostic is the expected action obtained from a softmax distribution over planner scores:
\begin{align}
\bar \xi_{\theta^{(m)}}(\mathsf{x})
=
\sum_{a\in\mathcal A}
\xi_a
\frac{\exp[\beta J_{\theta^{(m)}}(s(\mathsf{x}),a)]}
{\displaystyle\sum_{b\in\mathcal A}
\exp[\beta J_{\theta^{(m)}}(s(\mathsf{x}),b)]},
\label{eq:softmax-action-field}
\end{align}
where $\xi_a=(\delta_x,\delta_t)$ is the grid displacement associated with action $a$ and $\beta$ acts like an inverse temperature parameter that governs how spread out the distribution is. We average this field over the five agents and use it to visualize how the relative action scores determine an overall preferred direction of motion. For the geometric reconstruction, we first average the action scores themselves
\begin{align}
\overline J(\mathsf{x},a)=\frac{1}{5}\sum_{m=1}^{5}J_{\theta^{(m)}}(s(\mathsf{x}),a),
\end{align}
and then form the scalar softmax-average score
\begin{align}
\overline S(\mathsf{x})=
\sum_{a\in\mathcal A}\overline J(\mathsf{x},a)
\frac{\exp[\beta'\overline J(\mathsf{x},a)]}
{\displaystyle\sum_{b\in\mathcal A}\exp[\beta'\overline J(\mathsf{x},b)]},
\label{eq:softmax-score-field}
\end{align}
with $\beta' = 2$ giving the best results for the reconstruction procedure.

\begin{figure*}[t!]
\centering
\includegraphics[width=0.99\textwidth]{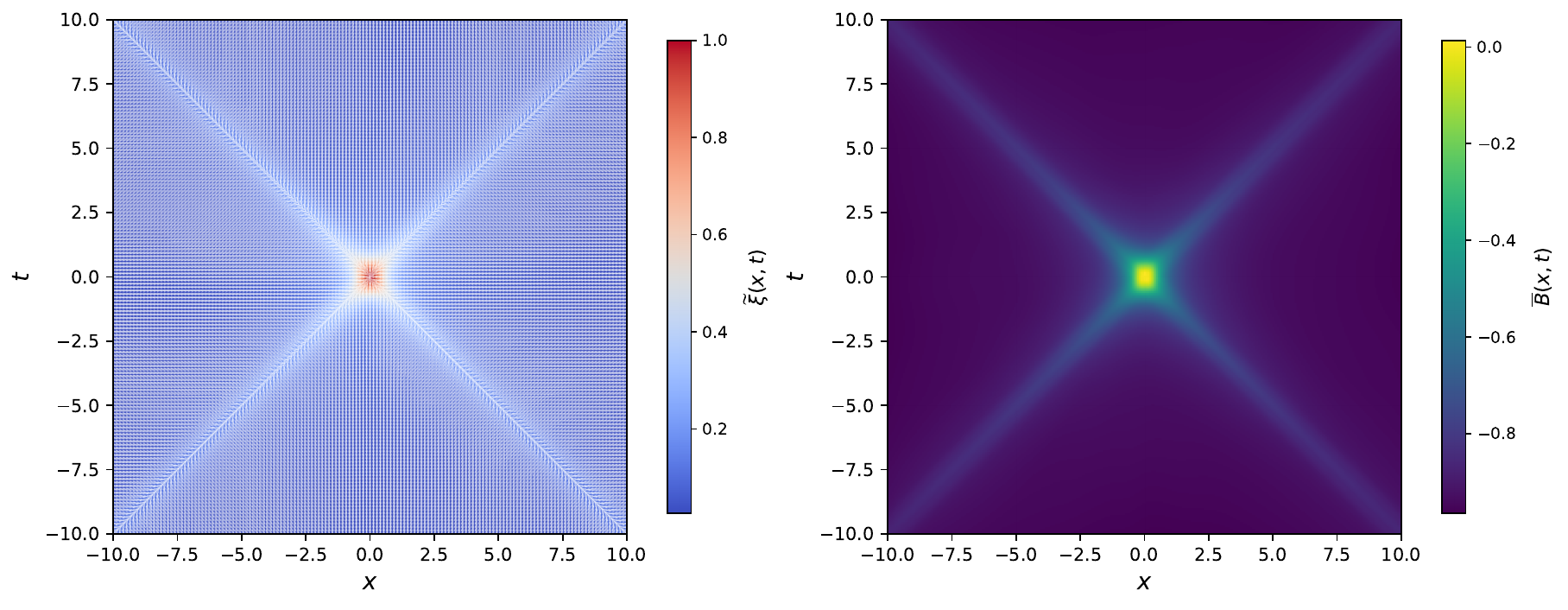}\caption{
Average planner memory output plots. Left: softmax expected-action field using $\beta=0.7$ and normalized arrows, with the choice of $\beta$ used only for visualization and not entering into training or planning. Right: averaged best planner score $\overline B$ at each detector placement. The trained planner maps the detector-generated reward landscape to a local action flow and a score surface that retain the underlying spacetime features (especially visible are the diagonal lightcone arms) inherited from the field correlations appearing in the quantum mutual information Eq. \eqref{eq:mutual-info}.
}
\label{fig:planner-diagnostics}
\end{figure*}

Figure~\ref{fig:planner-diagnostics} shows these fields after the training. The softmax expected-action field illustrates how the planner directs the RL agent towards higher-information regions along gradient-like flows according to the underlying spacetime structure present in Fig. \ref{fig:landscape}, which is most visible amongst the trajectories moving along or into the diagonal arms of the lightcone. The best-score field also displays these features and closely resembles Fig. \ref{fig:landscape} in structure, preserving both its broad central support and distinctive diagonal ridges. The best-score field additionally highlights that planning does alter the original landscape, most noticeably through thicker diagonal arms and widening diagonal edges. A likely explanation for the first is that wider favourable regions can receive high scores whenever one of the available actions leads towards them, while the second may arise from propagation of a weaker signal that makes it harder for the RL agent to differentiate the optimal path. However, the two plots put together provide us with an approximate representation of the spacetime structure.

\begin{figure*}[t!]
\centering
\includegraphics[width=1.02\textwidth, trim=7 5 10 3, clip]{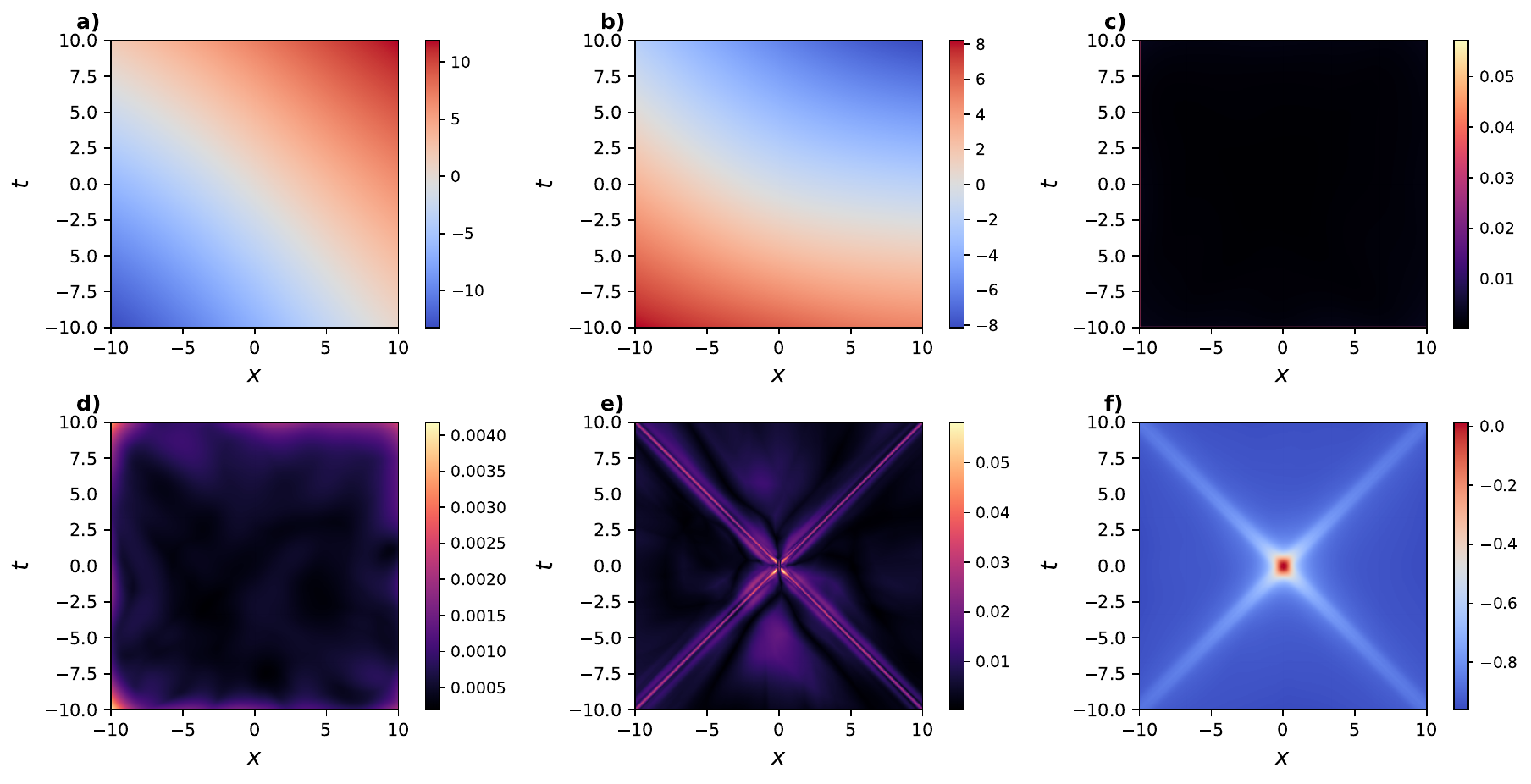}
\caption{
Internal diagnostics for the trained planner. Panels (a) and (b) show the first two principal components of the latent encoder; panel (c) shows the mean one-step latent dynamics error; panel (d) shows the current-state decoder reconstruction error; panel (e) shows the planner action gap; and panel (f) shows the planner-step reward on the selected action. Together, these quantities describe the latent representation, its one-step predictive dynamics, its connection to the observed detector-placement space, and the action preferences produced by the planner.
}
\label{fig:internal-diagnostics}
\end{figure*}

We can now inspect the quality of the internal model acquired by the RL agent. There are a few useful measures we can use to assess this quality, which we plot in Fig.~\ref{fig:internal-diagnostics}. The first two plots (a)---(b) show the two largest encoder principal components obtained from the encoded detector-grid observations, which indicate the two directions accounting for the greatest variation in the latent representation. Both have a clear boundary that they vary smoothly across over the detector grid, indicating that the latent representation retains spatial variation relevant to planning. The next plot (c) shows the mean one-step latent dynamics error between the predicted next latent state $\widehat z_{t+1}$ and encoded observed next state $z_{t+1}$, averaged over the available actions at each grid point, which we see remains essentially negligible across the explored region. The decoder reconstruction error in plot (d) is given by
\begin{align}
\varepsilon_{\rm dec}(\mathsf{x})
=
\left\|
D_{\theta_D}\!\left(E_{\theta_E}(o(\mathsf{x}))\right)
-o(\mathsf{x})
\right\|,
\end{align}
and shows that the latent states remain anchored to the observed detector-placement grid in the regions of interest, also demonstrating the poor edge representation and pointing to a root cause of bad transition model predictivity on edges. Plot (e) displays the planner action gap
\begin{align}
\Delta J_\theta(\mathsf{x})
=
J_\theta^{(1)}(\mathsf{x})-J_\theta^{(2)}(\mathsf{x}),
\end{align}
where $J_\theta^{(1)}(\mathsf{x})\geq J_\theta^{(2)}(\mathsf{x})$ are the two largest action scores at that grid point. This indicates that the best action to take along the lightcone branches is distinct relative to the next best, corroborating the story from Fig. \ref{fig:planner-diagnostics}. The planner selected-action reward in the final plot (f) exhibits the same lightcone-like support learned from the detector environment. Together, the plots in Fig. \ref{fig:internal-diagnostics} show that the RL agent has developed a predictively useful internal model whose latent representation, dynamics, and action preferences retain structured information about the detector environment. Further diagnostics are collected in Appendix~\ref{Appendix:InternalDiagnostics}.

Figures \ref{fig:planner-diagnostics} and \ref{fig:internal-diagnostics} clearly show the remnants of the spacetime structure exhibited in Fig. \ref{fig:landscape}. {We will now use the local structure of $\overline S$ to recover the causal geometry of the underlying spacetime, which in this case is flat $(1+1)$-dimensional Minkowski spacetime with metric tensor components $g_{tt}=-1$, $g_{tx}=g_{xt}=0$, $g_{xx}=1$, and eigenvalues $v_\pm=\pm1$. To do this, we will define a procedure to extract the relevant lightcone data $\{(\Delta t_i, \Delta x_i)\}_i$ and then use it to determine the coefficients of the quadratic form
\begin{align}
q(\Delta t,\Delta x) = g_{tt}\Delta t^2+2g_{tx}\Delta t\,\Delta x+ g_{xx}\Delta x^2,
\label{eq:planner-quadratic}
\end{align}
by solving for the null (lightlike) case $q(\Delta t, \Delta x) = 0$. 

In $(1+1)$ dimensions, two null directions determine the conformal class of a Lorentzian metric, and so writing $v=dx/dt$, the null condition becomes
\begin{align}
g_{tt}+2g_{tx}v+g_{xx}v^2=0.
\label{eq:null-slope-equation}
\end{align}
For lightcone branch slopes $v_+$ and $v_-$, their sum and product satisfy
\begin{align}
v_+ + v_- &= -\frac{2g_{tx}}{g_{xx}},\\
v_+v_- &= \frac{g_{tt}}{g_{xx}},
\label{eq:null-slope-metric-relations}
\end{align}
with conformal freedom in the choice of $g_{xx}$ that we choose to be $g_{xx}=1$, after which the lightcone branch geometry determines $g_{tt}$ and $g_{tx}$. For the extracted lightcone branch points $(\Delta t_i,\Delta x_i)$, these coefficients are obtained with equal weighting from the least-squares problem
\begin{align}
(\widehat g_{tt},\widehat g_{tx}) =\arg\min_{g_{tt},g_{tx}}\sum_i\left(g_{tt}\Delta t_i^2+2g_{tx}\Delta t_i\Delta x_i+\Delta x_i^2\right)^2.
\label{eq:planner-metric-fit}
\end{align}
We now need to find a good way of defining what points lie on the lightcone branches, as seen from the agent's memory. Since the planner assigns value according to locally accessible outcomes, the central high-reward region extends into neighbouring cells from which favourable states can be reached, so an absolute score threshold then selects much of this broad basin. The diagonal lightcone branches are more clearly characterized by their local prominence, which motivates comparing each planner score with the scores in its immediate neighbourhood; we will denote this 8-cell neighbourhood (excluding $\mf x$ itself) by $\mathcal N_1(\mathsf{x})$. We then construct the binary neighbour-comparison mask to be 
\begin{align}
M(\mathsf{x})=\mathbf 1\!\left[\frac{1}{8}\sum_{\mathbf y\in\mathcal N_1(\mathsf{x})}\mathbf 1_{\overline S(\mathsf{x})>\overline S(\mathbf y)}\geq 0.7\right],
\label{eq:neighbour-mask}
\end{align}
which thus retains cells whose scalar softmax-average score exceeds at least $70\%$ of the scores in their local neighbourhood, excluding cells with neighbourhoods outside the grid boundary. The idea is that this mask will strip away enough of the score-field features to leave just the central ridges of the lightcone branches, which can then be compared with the null directions of Eq. \eqref{eq:planner-quadratic}.

Branch extraction is then performed row by row on this binary mask, where the effective mask threshold is $M(\mathsf{x})>0.5$, so only cells with $M(\mathsf{x})=1$ are active, and the lightcone centre is fixed at $(x_c,t_c)=(0,0)$. For each retained time row $t_j$, the active cells are divided into the left and right regions $x<x_c$ and $x>x_c$, with rows sufficiently close to $t_c$ excluded from the fit. Within each side, the active cells are divided into contiguous runs, so if a run contains $N$ cells with mask values $p_k$, it is assigned the score
\begin{align}
S=\left(\sum_{k\in{\rm run}}p_k\right)\sqrt{N}.
\label{eq:branch-run-score}
\end{align}
Since the present mask is binary, this reduces to $S=N^{3/2}$, so the widest contiguous active run on each side is retained. A representative branch coordinate is then obtained from the peak cell of the selected run, and at most one left and one right branch point are retained for each time row. The upper and lower lightcone arms are both included through the positive and negative time rows, and the resulting left and right branch points are combined using
\begin{align}
\Delta t_i&=t_i-t_c,\\
\Delta x_i&=x_i-x_c,
\end{align}
which are then compared with the null directions of Eq. \eqref{eq:planner-quadratic} to compute the components of the metric tensor.

The results are shown in Fig. \ref{fig:neighbour-metric-fit}.

\begin{figure*}[t!]
\centering
\includegraphics[width=0.86\textwidth]{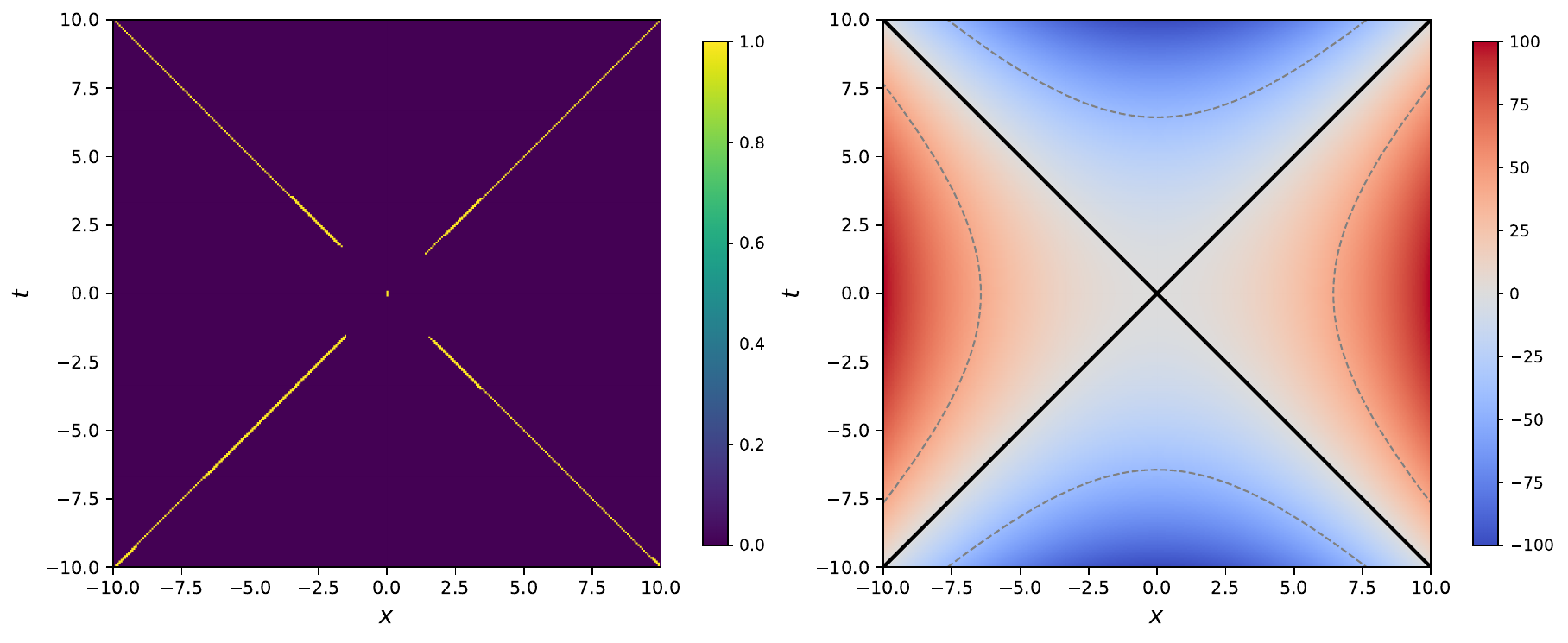}
\caption{
Reconstruction of the conformal metric from the scalar softmax-average planner score $\overline S$. Left: local neighbour-comparison mask obtained by retaining cells whose $\overline S$ value exceeds at least $70\%$ of the values in a radius-$1$ neighbourhood. Right: quadratic form displaying Lorentzian signature fitted to the extracted lightcone branches with the conformal convention $g_{xx}=1$. The resulting null directions closely reproduce those of conformally flat $(1+1)$-dimensional Minkowski spacetime.
}
\label{fig:neighbour-metric-fit}
\end{figure*}

In the coordinate ordering $(t,x)$, the resulting coefficients obtained from the five-agent averaged field are assembled into the metric representative
\begin{align}
g_{\rm fit}
=
\begin{pmatrix}
-0.999878 & 7.787\times10^{-8}\\
7.787\times10^{-8} & 1
\end{pmatrix},
\label{eq:fit-metric}
\end{align}
whose eigenvalues are approximately $(-0.999878,1)$ with determinant $-0.999878$ and corresponding null slopes $v_\pm\simeq(-0.999939,0.999939)$, thus yielding a metric with Lorentzian signature. For comparison, flat $(1+1)$-dimensional Minkowski spacetime has $g_{tt}=-1$, $g_{tx}=0$, $g_{xx}=1$, and $v_\pm=\pm1$. The reconstructed time component differs from its exact value by approximately $1.22\times10^{-4}$, the fitted off-diagonal contribution is negligible relative to the diagonal entries, and the null slopes differ from $\pm1$ by approximately $6.10\times10^{-5}$ \footnote{These values characterize the fit to the averaged planner field and should not be confused with a statistical uncertainty on the metric reconstruction. Estimating such an uncertainty would require, for example, fitting the agents separately or resampling the extracted lightcone branches.}. Within the conformal convention $g_{xx}=1$, the result therefore recovers a representative of the conformal class that closely approximates the Minkowski metric in these coordinates, demonstrating that the learned planner representation retains sufficient geometric information to approximately reconstruct the causal structure of the underlying spacetime.

\section{Discussion and Outlook} \label{Section:Discussion}


The results from Section \ref{Section:AgentMemory} provide a clear picture where an RL agent given a quantum information task embedded in a spacetime learns to accomplish that task while adhering to the geometry of the underlying spacetime. Through construction of a rich internal model of the environment that contains enough geometric and causal information, the agent is able to quantitatively reconstruct the Lorentzian metric of spacetime. Not only is the spacetime information implicitly hidden in the RL agent's environment feedback----in addition to the fact that the RL agent is initialized with no information about what spacetime is----but the architecture of the RL agent also scrambles that information into an internal latent representation, disconnected with the exterior spacetime grid. This connects the present construction to work on \emph{intuitive physics}, where agents can acquire internal representations supporting physical prediction without explicitly evaluating any equations of motion~\cite{Battaglia2013,Lake2017,Piloto2022}. In particular, predictive models have been shown to acquire identifiable physical concepts from experience despite the absence of them in the explicit output of the prediction task~\cite{Piloto2022}.

This suggests an operational perspective on \emph{spacetime concept formation}, where an RL agent does not need to possess an explicit symbolic representation of a Lorentzian manifold in order for spacetime structure to organize its behaviour. Instead, regularities in its interactions can become encoded as relations among internal states, predictions, and actions. In the present model the planner never explicitly evaluates $ g_{\mu\nu}\Delta x^\mu\Delta x^\nu = 0$ when choosing an action, yet the memory supporting those choices becomes sensitive to operational regularities from which this null condition can subsequently be recovered. In this restricted sense, the learned representation provides a primitive model of a spacetime concept as an action-relevant organization of experience that implicitly captures causal relations, analogous to the metric tensor. This is consistent with the view that memory goes beyond functioning as just an inert storage device, but plays a more active role as a resource for generating future behaviour ~\cite{BriegelDeLasCuevas2012,LeMaitreKrumm2024,briegel2025projective}.

This point of view does highlight an important distinction however, which is that the RL agent does not carry the actual metric tensor object in its memory. As we saw earlier, it's memory acquires a geometric flavour from learning, and the metric tensor is subsequently extracted using external tools supplied when studying it, such as lightcone masking and conformal gauge fixing. While using such geometrically informed post-processing procedures is unsatisfactory compared to the larger vision of explicitly emergent causal or metric structure in the RL agent's memory, we view this work as merely a first step towards this setting. A stronger realization of this programme could of course make geometric inference itself part of the learned world model, for instance as a learned sub-network of the RL agent's internal model. 

Minkowski spacetime also provides a deliberately simple test of the above ideas that gives us increased clarity by making every step of the reconstruction process inspectable, especially the memory of the RL agent. Of course its recovery --- particularly in the $(1+1)$-dimensional case --- is not in and of itself surprising, since the causal structure is known analytically and is already reflected in the known field correlations from which the detector data are generated. One should realize however, that the architecture and problem setup are agnostic to the underlying geometry of the data, so in the general setting they can still be employed to learn the structure even when it is not known in advance, meaning that the same strategy can be applied when detector generated quantum information is available but the spacetime from which it originates is not. More robust tests of the reconstruction under coordinate changes and different detector smearings should be undertaken as well before making this step. This would substantially strengthen the evidence that the learned relations represent spacetime structure as opposed to contingent properties of the chosen coordinates or probes.

The question of how far we can take this interpretation will require several extensions. First, quantum mutual information is not a pure measure of signalling as it contains both causally mediated and state-dependent correlations, including correlations at spacelike separation. Future agents should thus receive several complementary operational quantities that can tease these differences apart, such as signalling estimators or spin correlations, and more generally observables that are separately sensitive to commutator and anticommutator contributions. These may then allow for causal propagation and scale information to be distinguished and could provide the additional operational information required to reconstruct more than just the conformal structure. The setting should likewise be extended to $(3+1)$-dimensional and curved spacetimes, where the local causal structure varies across the environment and cannot be identified with two globally straight lightcone branches like in Minkowski.

Finally, one can consider more challenging or practical quantum information tasks as the learning objective, since maximizing a scalar function of mutual information provides a nice transparent proof of principle, but a more compelling test would place the RL agent in a task for which learning spacetime structure is actually necessary for completing the task. Quantum communication or adaptive entanglement distribution through detector networks could be natural and interesting possibilities, which may also require the RL agent to utilize longer-horizon planning that would test how it progresses from implicitly exploiting causal regularities to constructing increasingly general representations of the spacetime in which it acts. The quantum information task studied here shows how the causal structure of Minkowski spacetime can emerge in the memory of an agent, opening the doors to the more ambitious question of whether agents can autonomously discover an explicit spacetime representation from operational experience alone. 

\acknowledgments
The authors would like to thank Hans J. Briegel for initial discussions leading to the conception of the project, as well as for ongoing discussions and feedback on drafts. This research was funded in whole or in part by the European Union (ERC Advanced Grant, QuantAI, No. 101055129) and the Austrian Science Fund (FWF) [SFB BeyondC F7102, 10.55776/F7102]. The views and opinions expressed in this article are however those of the author(s) only and do not necessarily reflect those of the European Union or the European Research Council --- neither the European Union nor the granting authority can be held responsible for them. For open access purposes, the authors have applied a CC BY public copyright license to any author-accepted manuscript version arising from this submission. T. R. Perche is hosted at the Physics Department of Stockholm University through a Wenner-Gren Fellowship and is thankful for partial financial support from the Olle Engkvist Foundation (no.225-0062).

The authors would also like to disclose the use of AI tools (OpenAI Codex with GPT-5.5) for the creation of Fig. \ref{fig:agent-architecture} as well as code generation for obtaining plots, optimizing and debugging code, and automation of standard and publicly available code blocks. All ideas, results, and scientific conclusions originate with the authors. We have verified everything thoroughly and take full responsibility for everything contained in the manuscript. 

\bibliography{RQML}

\appendix
\onecolumngrid
\section*{Appendices}
\phantomsection\label{Section:Appendices}

\section{Supplementary Internal Diagnostics} \label{Appendix:InternalDiagnostics}

Figure \ref{fig:internal-diagnostics} in Section \ref{Section:AgentMemory} displayed some of the important diagnostics for assessing the quality of the RL agent's internal model, and here we will provide more complementary diagnostics to those plots. These remaining plots shown in Fig.~\ref{fig:internal-diagnostics-full} provide a more detailed look at the properties of the model components, and their role is to solidify that the score generated by the planner, which is used for the metric extraction, is produced by properly functioning internal predictive models along with the encoder/decoder.

\begin{figure*}[t!]
    \centering
    \includegraphics[width=0.90\textwidth]{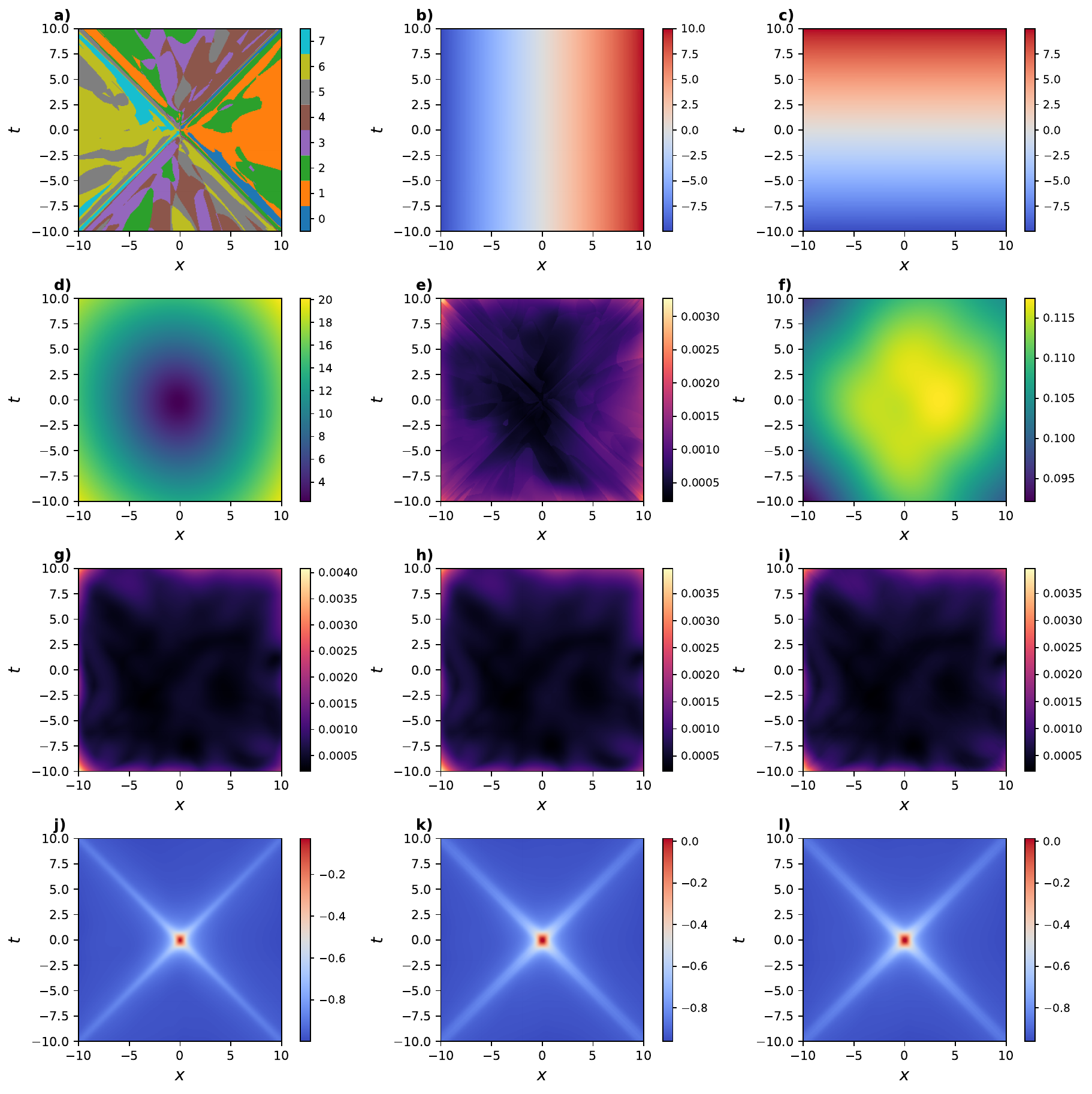}
    \caption{
    Supplementary internal diagnostics for the trained planner.  Plot (a) shows the argmax action splitting into each action direction; plots (b) and (c) show the decoded coordinates; plot (d) shows the latent norm; plot (e) shows the dynamics error; plot (f) shows the mean predicted latent step norm; plots (g)---(i) show next-state reconstruction errors; and plots (j)---(l) show reward model projections.
    }
    \label{fig:internal-diagnostics-full}
\end{figure*}

Plot (a) displays the argmax action map and records how the planner partitions the grid into local action regions, clearly displaying the lightcone structure on the diagonals but also drawing attention to different local action domains. Plots (b) and (c) give the decoded coordinates which unambiguously show that the $x$ and $t$ coordinate organization is preserved by the encoder and decoder. Plot (d) gives the latent norm, which remains finite and spatially structured over the grid, and indicates that the latent state components are better shaped as we move closer to the goal state. Plots (e) and (f) display the action-conditioned dynamics error and the mean predicted latent step norm, both of which have the lightcone structure somewhat visible in them, with the first indicating that action-conditioned dynamics becomes slightly less reliable the further away from the goal that we move, and the second showing that the imagined latent displacement across the grid remains fairly controlled. Plots (g)---(i) show that the next-state reconstruction errors remain small away from the extreme boundary region. Finally, plots (j)---(l) compare three reward model projections: the mean over actions, the maximum over actions, and the reward assigned to the planner-selected action. Their common lightcone-shaped support is consistent with the reward range reported in Sec.~\ref{Section:AgentMemory} and with the planner-step reward shown in Fig.~\ref{fig:internal-diagnostics}. Thus, all plots are consistent with and support the narrative that the RL agent learned a useful representation of the spacetime to recover the Lorentzian quadratic form in Eq.~\eqref{eq:fit-metric}.

\end{document}